\documentclass[a4paper,10pt,twoside]{cpc-hepnp}

\usepackage{multicol}
\usepackage{graphicx}
\usepackage{booktabs}
\usepackage{amssymb,bm,mathrsfs,bbm,amscd}
\usepackage[tbtags]{amsmath}
\usepackage{lastpage}

\begin{document}
%\begin{CJK*}{GBK}{song}

%\fancyhead[c]{\small Chinese Physics C~~~Vol. 37, No. 1 (2013)
%010201} \fancyfoot[C]{\small 010201-\thepage}

%\footnotetext[0]{Received 14 March 2009}

\title{Woods-Saxon-Gaussian Potential and Alpha-Cluster Structures of Alpha+Closed Shell Nuclei\thanks{ This work is supported by the National Natural Science Foundation of China (Grant No.~11535004, 11761161001, 11375086, 11120101005, 11175085 and 11235001), by the National Key R\&D Program of China (Contract No.~2018YFA0404403, 2016YFE0129300), and by the Science and Technology Development Fund of Macau under Grant No.~008/2017/AFJ.
}}

\author{%
      Dong Bai$^{1}$\email{dbai@itp.ac.cn}%
\quad and \quad Zhongzhou Ren$^{2}$\email{Corresponding Author: zren@tongji.edu.cn}%
%\quad LI Da-Ming$^{1}$
%\quad F. Jone$^{2}$
}
\maketitle

\address{%
$^1$ School of Physics, Nanjing University, Nanjing, 210093, China\\
$^2$ School of Physics Science and Engineering, Tongji University, Shanghai, 200092, China\\
}

\begin{abstract}
The Woods-Saxon-Gaussian (WSG) potential is proposed as a new phenomenological potential to describe systematically the level scheme, electromagnetic transitions, and alpha-decay half-lives of the alpha-cluster structures in various alpha+closed shell nuclei.
% $^{212}\text{Po}={}^{208}\text{Pb}+\alpha$. 
 It modifies the original Woods-Saxon (WS) potential with a shifted Gaussian factor centered at the nuclear surface. We determine the free parameters in the WSG potential by reproducing the correct level scheme of $^{212}\text{Po}={}^{208}\text{Pb}+\alpha$. It is found that the resulting WSG potential matches with the M3Y double-folding potential at the surface region and makes corrections to the inner part of the cluster-core potential. We also find that the WSG potential with the almost same parameters determined for $^{212}$Po (except for a rescaled radius) could also be used to describe alpha-cluster structures in $^{20}\text{Ne}={}^{16}\text{O}+\alpha$ and $^{44}\text{Ti}={}^{40}\text{Ca}+\alpha$. In all three cases, the calculated values of the level schemes, electromagnetic transitions, and alpha-decay half-lives agree with the experimental data, which shows that the WSG potential could indeed grasp many important features of the alpha-cluster structures in alpha+closed shell nuclei. The study here is a useful complement to the existing cluster-core potentials in literature. The Gaussian form factor centered at the nuclear surface might also help deepen our understanding on the alpha-cluster formation taking place around the same place.
\end{abstract}

%
%\begin{pacs}
%1--3 PACS(Physics and Astronomy Classification Scheme, http://www.aip.org/pacs/pacs.html/)
%\end{pacs}

%\footnotetext[0]{\hspace*{-3mm}\raisebox{0.3ex}{$\scriptstyle\copyright$}2013
%Chinese Physical Society and the Institute of High Energy Physics
%of the Chinese Academy of Sciences and the Institute
%of Modern Physics of the Chinese Academy of Sciences and IOP Publishing Ltd}%

\begin{multicols}{2}

\section{Introduction}

Alpha clusters play a crucial role in nuclear physics. Alpha clustering in light nuclei was proposed in the 1930s \cite{Hafstad:1938} and has been studied intensively since then, especially after the proposal of the existence of alpha-particle condensates in Hoyle and Hoyle-like states in 2001 \cite{Tohsaki:2001an}. See, e.g. , Ref.~\cite{Bai:2018gqt} for a recent discussion on the impact of repulsive four-body interactions of alpha particles on physical properties of alpha-particle condensates. Inspired by Ref.~\cite{Tohsaki:2001an}, nonlocalized clustering is proposed as a new concept of cluster physics, and an exemplifying calculation is carried out for the inversion-doublet band of $^{20}$Ne \cite{Zhou:2013ala}. Alpha clustering could also exist in heavy/superheavy nuclei, as suggested by the observation of alpha decay in these nuclei. The theoretical explanation of alpha decay dates back to the celebrating papers by Gamow, Gurney, and Condon \cite{Gamow:1928,Gurney:1928}. Nowadays, lots of phenomenological models have been proposed to explain the observed alpha-decay and spectroscopic data, in which the parent nucleus is modeled as a binary system made of a tightly-bound alpha cluster and the remaining core nucleus (see, e.g., Ref.~\cite{Buck:1990zz,Buck:1992zz,Buck:1995zza,Buck:1996zza,Denisov:2009zza,Ibrahim:2010zz,Delion:2015wro,Denisov:2015wka,Mohr:2017zot,Delion:2018rrl,Ismail:2018cdy}). The relative motion between these two constituents then gives rise to various properties of parent nuclei. For the later convenience, we shall call these phenomenological models as binary cluster models (BCMs). It is interesting to note that some of BCMs could describe alpha clustering not only in heavy nuclei but also in light and medium-mass nuclei such as $^{20}$Ne and $^{44}$Ti \cite{Buck:1995zza}. In Ref.~\cite{Xu:2005ukj,Ni:2009vzd,Ni:2009zza,Ni:2009zzb,Ni:2009zz,Ni:2010zza,Ni:2010zzc}, a new phenomenological model named density-dependent cluster model is proposed and developed, which takes into consideration the impact of nuclear density distributions and gives better descriptions of the alpha-decay data. Microscopic descriptions of alpha decay and alpha clustering in heavy nuclei are also pursued. One of the milestones is the celebrated work done by Ref.~\cite{Varga:1992zz}, which gives a satisfying description of the ground-state alpha decay in $^{212}$Po based on the cluster-configuration shell model. For recent works on microscopic descriptions of $^{212}$Po, see, e.g., Ref.~\cite{Astier:2009bs,Astier:2010sn,Delion:2012zz,Ropke:2014wsa,Xu:2015pvv}. We also recommend Ref.~\cite{Delion:2010} for a comprehensive introduction to alpha clustering in heavy nuclei.

%Alpha clusters play an important role in both light and heavy/superheavy nuclei, and various phenomenological models have been proposed to provide systematic descriptions for these alpha-cluster structures \cite{Gamow:1928,Gurney:1928,Buck:1990zz,Buck:1992zz,Buck:1995zza,Buck:1996zza,Yamada:2003cz,Delion:2010,Delion:2015wro,Delion:2018rrl,Xu:2005ukj,Ni:2009vzd,Ni:2009zza,Ni:2009zzb,Ni:2009zz,Ni:2010zza,Ni:2010zzc,Denisov:2005ax,Denisov:2009ng,Denisov:2009zza,Denisov:2015wka,Ismail:2013bma,Ibrahim:2010zz,Wang:2013gk,Mohr:2006qm,Mohr:2016ofn,Mohr:2017zot,Bai:2018gqt,Ismail:2018cdy,Seif:2017gtk,Duppen:2018weh}. Suggested by the observed alpha decays and spectroscopic data, many of light and heavy/superheavy nuclei could be viewed as binary systems made of a tightly-bound alpha cluster and the remaining core nucleus. The relative motion of these two constituents then gives rise to various properties of light and heavy/superheavy nuclei. For the later convenience, we shall call these phenomenological models as binary cluster model (BCM).

A successful application of BCM depends crucially on the choice of the cluster-core nuclear potential. The simplest choice might be the square-well potential \cite{Buck:1990zz}. More realistic choices for the phenomenological potential include the double-folding potential based on, e.g., the M3Y nucleon-nucleon interaction \cite{Xu:2005ukj}, the so-called Cosh potential obtained by fitting the double-folding potential \cite{Buck:1992zz}, the Woods-Saxon (WS) potential inspired by shell model \cite{Ni:2009vzd}, the WS$+$WS$^3$ potential \cite{Buck:1995zza,Ibrahim:2010zz}, a new local potential proposed in Ref.~\cite{Wang:2013gk}, etc. With the help of the Bohr-Sommerfeld-Wildermuth quantization condition \cite{Wildermuth:1977}
%,Buck:1990zz,Buck:1992zz,Buck:1993sku,Buck:1996zza,Buck:1994zz,Buck:1995zza,Buck:1995zz,Xu:2005ukj,Ni:2009vzd,Ni:2009zza
 to mimic the Pauli blocking effects, typically all these potentials could reproduce quite well alpha-decay half-lives for heavy and superheavy alpha-emitters. However, further studies show that, the M3Y potential, the Cosh potential, and the WS potential fail to explain the observed energy spectrum of $^{212}$Po$={}^{208}\text{Pb}+\alpha$, which is a canonical example of alpha+closed shell nuclei, and give typically inverted level schemes \cite{Buck:1995zza,Buck:1996zza}. This makes the WS$+$WS$^3$ potential almost the unique choice for the spectroscopic studies of alpha-cluster structures of $^{212}$Po in literature. In this work, we would like to propose a new phenomenological potential between the cluster and core nuclei. $^{212}$Po is used as a benchmark nucleus to determine the free parameters in the potential. Later, these parameters with radii being rescaled are shown to be able to describe alpha clusters in light and medium-weight nuclei as well.

\section{Formalism}

In general, a cluster-core nuclear potential, which aims to provide a satisfying description for alpha decays of $^{212}$Po, should match with the WS potential or the M3Y potential at the surface region. However, as noted above, the WS potential alone fails to give the correct level scheme. Also, it is well-known that the alpha-cluster formation process actually takes place around the nuclear surface \cite{Delion:2010}. Within such picture, the alpha cluster loses its identity and merges with the core nucleus when moving towards the center of the core nucleus. A satisfying phenomenological potential might be able to give some hints on this process. With these in mind, we introduce the following general form for the cluster-core nuclear potential,
\begin{align}
V_N(r)=-\frac{V_0\,f(r)}{1+\exp\left[{(r-R)}/{a}\right]},
\end{align}
with $f(r)$ being the form factor that encodes corrections to the inner part of the WS potential. Microscopically, this nontrivial form factor $f(r)$ might be related to the alpha-cluster formation around the nuclear surface and other quantum mechanical effects. The form factor $f(r)$ could be studied by using the Gaussian basis,
\begin{align}
f(r)=1+\sum_i\alpha_i\exp[-\beta_i(r-R_i)^2],
\end{align}
thanks to the over-completeness of the Gaussian basis \cite{Hiyama:2003cu}. In this work, for simplicity, we take only the leading-order correction,
\begin{align}
V_N(r)=-\frac{V_0}{1+\exp\left[{(r-R)}/{a}\right]}\left\{1+\alpha\exp[-\beta(r-R)^2]\right\}.
\label{WSG}
\end{align}
Here, it is crucial to note that the same $R$ in the WS potential is used as $R_1$ in the Gaussian form factor. In other words, we plan to modify the WS potential with a Gaussian form factor center at the nuclear surface. This is by no means trivial. Generally, an arbitrary function $f(r)$ does not allow such an identification. As shown later, this choice is able to give reasonable descriptions on alpha-cluster structures in alpha+closed shell nuclei. It is natural to guess that the underlying physics is related to the alpha-cluster formation process mentioned before.In fact, the idea to use additional Gaussian alpha-like components was used first within the cluster-configuration shell model by Ref.~\cite{Varga:1992zz} and later in Ref.~\cite{Delion:2012zz}.  In the following, we shall call Eq.~\eqref{WSG} as the Woods-Saxon-Gaussian (WSG) potential. We intend to use the WSG potential to provide a systematic treatment of various nuclear structural and alpha-decay properties. $V_0$, $R$, $a$, $\alpha$, and $\beta$ are five free parameters to be determined by reproducing some experimental observations. The usefulness of the WSG potential will be justified by a comparison of the theoretical calculations with the experimental data on $^{212}$Po. The WSG potential in Eq.~\eqref{WSG} could certainly be improved further by including more Gaussian terms in the braces.  

Given values of the free parameters, the WSG potential can be used to calculated the spectrum of $^{212}$Po. Here, we use the Bohr-Sommerfeld-Wildermuth quantization condition
\begin{align}
&\int_{r_1}^{r_2}\sqrt{\frac{2\mu}{\hbar}[E_L-V(r)]}\mathrm{d}r=(G-L+1)\frac{\pi}{2},\\
&V(r)=V_N(r)+V_C(r)+V_L(r),
\end{align}
to calculate the spectrum. $V_N(r)$ is the WSG potential given by Eq.~\eqref{WSG}. $V_C(r)$ is the Coulomb potential defined as
\begin{equation}
V_C(r)=
\begin{cases}
\frac{Z_cZ_\alpha e^2}{r},\quad\quad &r\geq R,\\
\frac{Z_cZ_\alpha e^2}{2R}\left[3-\left(\frac{r}{R}\right)^2\right], \quad\quad & r<R.
\end{cases}
\end{equation} 
$V_L(r)$ is the centrifugal potential in the Langer approximation
\begin{align}
V_L(r)=\frac{\hbar^2}{2\mu r^2}\left(L+\frac{1}{2}\right)^2.
\end{align}
The quantum number $G$ is chosen to be $G=18$ for $^{212}$Po \cite{Buck:1995zza,Buck:1996zza}. $r_1$ and $r_2$ are the inner classical turning points. $\mu$ is the reduced mass of alpha cluster. $E_L$ is the energy of the state with the orbital angular momentum $L$. 

The WSG potential can also be used to study alpha decays and electromagnetic transitions of $^{212}$Po. The alpha-decay width is estimated by using the two-potential approach \cite{Gurvitz:1987,Gurvitz:1988},
\begin{align}
&\Gamma_\alpha=P_\alpha F\frac{\hbar^2}{4\mu}\exp\left[-2\int_{r_2}^{r_3}\mathrm{d}rk(r)\right],\nonumber\\
&k(r)=\sqrt{\frac{2\mu}{\hbar^2}\left|E_L^\text{Exp}-V(r)\right|},\nonumber\\
&F\int_{r_1}^{r_2}\mathrm{d}r\frac{1}{k(r)}\cos^2\left(\int_{r_1}^{r}\mathrm{d}r'k(r')-\frac{\pi}{4}\right)=1,
\label{AlphaDecay}
\end{align}
with $P_\alpha$ being the preformation probability of the alpha cluster in $^{212}$Po and $r_3$ being the outer classical turning points of the Coulomb barrier. The reduced quadrupole transition strength from an initial state of the angular momentum $L$ to a final state of the angular momentum $L-2$ could be calculated by \cite{Buck:1996zza}
\begin{align}
%&\Gamma_\gamma(\text{E}2;L_i\to L_f)=\frac{12\pi e^2}{225}(1+\alpha_T)\left(\frac{E_\gamma}{\hbar c}\right)^5 B(\text{E}2;L_i\to L_f),\\
&B(\text{E}2;L\to L-2)=\frac{15\beta^2_2}{8\pi}\frac{L(L-1)}{(2L+1)(2L-1)}\nonumber\\
&\qquad\qquad\qquad\times\left|\int_0^\infty\psi_{L-2}(r)^*r^2\psi_{L}(r)\mathrm{d}r\right|^2,\\
&\beta_2=e\frac{Z_cA_\alpha^2+Z_\alpha A_c^2}{(A_c+A_\alpha)^2},
\end{align} 
with %$\alpha_T$ being the internal conversion coefficient, $E_\gamma$ being the transition energy, and 
$\psi_L(r)$ being the radial wave function of the angular momentum $L$. 
%The total half-life $T_{1/2}$ and the alpha-decay branching ratio are then obtained by
%\begin{align}
%&T_{1/2}=\frac{\hbar \ln 2}{\Gamma_\alpha+\Gamma_\gamma},\\
%&b_{\alpha}=\frac{\Gamma_\alpha}{\Gamma_\alpha+\Gamma_\gamma}.
%\end{align}

\section{Numerical Results}

In the rest part of this article, we shall first apply the WSG potential to study alpha-cluster structures in $^{212}$Po. The free parameters of the WSG potential are determined by minimizing $\chi^2=\sum_L(E_L^\text{Exp}-E_L^\text{Present})^2$ and reproduce simultaneously the correct level scheme of $^{212}$Po. Experimentally, the energy of the $16^+$ state has not been measured yet, and it is reasonable to assume that $E_{14}^\text{Present}<E_{18}^\text{Present}<E_{16}^\text{Present}$, which is able to hinder the electromagnetic decay of the $18^+$ state. As a result, the ${18}^+$ state would mainly decay through alpha decay, whose decay rate is then strongly suppressed by the large centrifugal barrier. This could provide a natural explanation of the isomeric character of the $18^{+}$ state and its observed decaying branching ratio monopolized by alpha decay ($\sim 100\%$). Such an assumption is also consistent with previous studies in, e.g., Ref.~\cite{Ibrahim:2010zz}. After a fair amount of trial and error, the free parameters in the WSG potential are chosen to be
\begin{align}
&V_0=203.3 \text{ MeV}, \quad a=0.73 \text{ fm}, \quad R=6.73 \text{ fm},\nonumber\\
&\alpha=-0.478, \quad \beta=0.054 \text{ fm}^{-2}.
\label{WSGParameters}
\end{align}

In Fig.~\ref{WSGvsM3Y}, we compare the WSG potential with the M3Y potential obtained by double-folding the M3Y nucleon-nucleon interactions. Technical details for calculating the M3Y double-folding potential could be found in Appendix \ref{MSA}. The WSG potential is found to match approximately with the M3Y potential in the surface region with $r>6$ fm. 

\begin{center}
\includegraphics[width=0.45\textwidth]{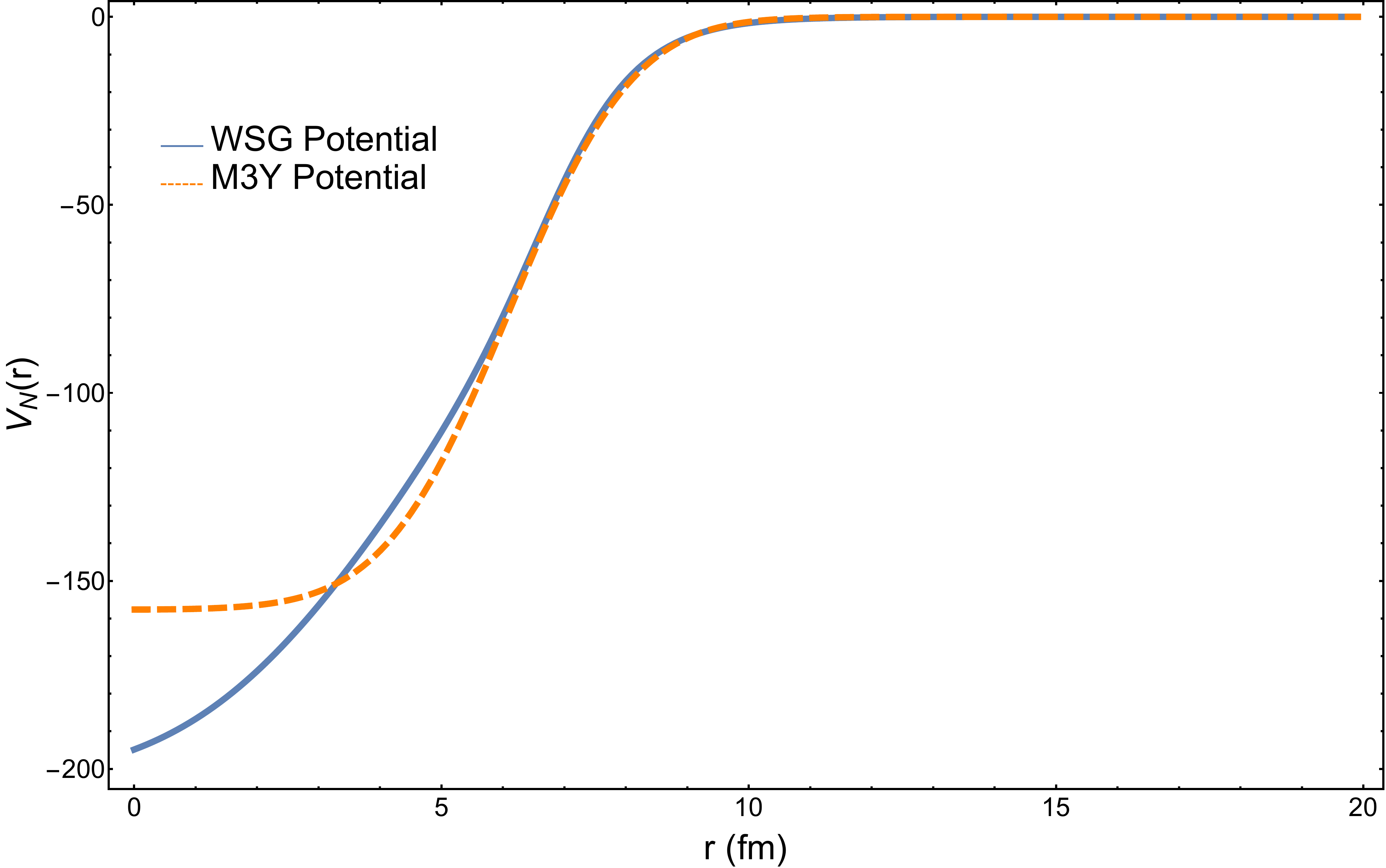}
\figcaption{\label{WSGvsM3Y} 
The WSG potential with the parameters given by Eq.~\eqref{WSGParameters} and the M3Y potential with the renormalization factor $\lambda=0.51$. The WSG potential matches approximately with the M3Y potential in the surface region with $r>6$ fm.}
\end{center}

In Table \ref{ExpvsTheor}, we tabulate the experimental level scheme of $^{212}$Po with theoretical results obtained by using the WSG potential with the parameters given by Eq.~\eqref{WSGParameters}. The reduced transition strengths $B(\text{E}2\!\downarrow)$ obtained by using the WSG potential are tabulated in Table \ref{BE2}. All values therein are given in the unit of the Weisskopf unit (W.u.). The theoretical values agree with the experimental ones within a factor of 2, and reproduce the trend nicely. The alpha-decay half-lives $T_{1/2}^{\alpha}=\hbar\ln2/\Gamma_\alpha$ are calculated using Eq.~\eqref{AlphaDecay}, and are tabulated in Table \ref{AlphaDecayTable}. The preformation factor is taken to be $P_\alpha=0.6$ to achieve better agreement with experimental data. It turns out that most of the theoretical values are within a factor of 2 compared to the experimental values.

%We also reproduce the corresponding results of Ref.~\cite{Ibrahim:2010zz} by Ibrahim \emph{et al.} in Table \ref{ExpvsTheor} and \ref{BE2}. Ref.~\cite{Ibrahim:2010zz} determines free parameters of the WS$+$WS$^3$ potential by using a hybrid method with various parameters determined by the asymptotic form of the microscopic M3Y potential. It is straightforward to see that the results obtained by the WSG potential are numerically close to those in Ref.~\cite{Ibrahim:2010zz}, which is quite encouraging. 

%\begin{figure}[tb]
%\centering
%\end{figure}
%
%\begin{table}
\begin{center}
\tabcaption{The experimental level scheme of $^{212}$Po and the theoretical calculations obtained with the WSG potential with the parameters given by Eq.~\eqref{WSGParameters}. The experimental values are taken from Ref.~\cite{Astier:2009bs,Astier:2010sn}.\label{ExpvsTheor}
}. %We also reproduce the results from Ref.~\cite{Ibrahim:2010zz} for comparison.
%}
\footnotesize
%\begin{center}
\begin{tabular}{cccc}
\hline
\hline
\hspace{2mm}$J^\pi$\hspace{5mm} & \hspace{5mm}{$E_L^\text{Exp}$ (MeV)}\hspace{5mm} & \hspace{10mm}{$E_L^\text{Present}$ (MeV)}
%\hspace{10mm} & \hspace{10mm}{$E_L^\text{Ibrahim}$ (MeV)}
\hspace{2mm}\\[0.5ex]  
\hline
\hspace{5mm}$0^+$\hspace{10mm} &  \hspace{10mm}{0.000}\hspace{10mm} & \hspace{10mm}{0.076}
%\hspace{10mm} & \hspace{10mm}{-0.060}
\hspace{5mm} \\
\hspace{5mm}$2^+$\hspace{10mm} &  \hspace{10mm}{0.727}\hspace{10mm} & \hspace{10mm}{0.255}
%\hspace{10mm} & \hspace{10mm}{0.131}
\hspace{5mm} \\
\hspace{5mm}$4^+$\hspace{10mm} &  \hspace{10mm}{1.132}\hspace{10mm} & \hspace{10mm}{0.596}
%\hspace{10mm} & \hspace{10mm}{0.478}
\hspace{5mm} \\
\hspace{5mm}$6^+$\hspace{10mm} &  \hspace{10mm}{1.355}\hspace{10mm} & \hspace{10mm}{1.050}
%\hspace{10mm} & \hspace{10mm}{0.928}
\hspace{5mm} \\
\hspace{5mm}$8^+$\hspace{10mm} &  \hspace{10mm}{1.476}\hspace{10mm} & \hspace{10mm}{1.581}
%\hspace{10mm} & \hspace{10mm}{1.442}
\hspace{5mm} \\
\hspace{5mm}$10^+$\hspace{10mm} & \hspace{10mm}{1.834}\hspace{10mm} & \hspace{10mm}{2.141}
%\hspace{10mm} & \hspace{10mm}{1.975}
\hspace{5mm} \\
\hspace{5mm}$12^+$\hspace{10mm} & \hspace{10mm}{2.702}\hspace{10mm} & \hspace{10mm}{2.676}
%\hspace{10mm} & \hspace{10mm}{2.477}
\hspace{5mm} \\
\hspace{5mm}$14^+$\hspace{10mm} & \hspace{10mm}{2.885}\hspace{10mm} & \hspace{10mm}{3.108}
%\hspace{10mm} & \hspace{10mm}{2.888}
\hspace{5mm} \\
\hspace{5mm}$16^+$\hspace{10mm} & \hspace{10mm}{$-$}\hspace{10mm} & \hspace{10mm}{3.328}
%\hspace{10mm} & \hspace{10mm}{3.126}
\hspace{5mm} \\
\hspace{5mm}$18^+$\hspace{10mm} & \hspace{10mm}{2.921}\hspace{10mm} & \hspace{10mm}{3.152}
%\hspace{10mm} & \hspace{10mm}{3.066}
\hspace{5mm} \\
 \hline
\hline
\end{tabular}
%\end{center}
%\end{table}
\end{center}
%
%\begin{table}
\begin{center}
\tabcaption{Theoretical values of the reduced quadrupole transition strengths of $^{212}$Po obtained with the WSG potential with the parameters given by Eq.~\eqref{WSGParameters}. 
%The experimental values are taken from Ref.~\cite{Buck:1996zza}. 
All values are in the unit of the so-called Weisskopf unit (W.u.) given by 1 W.u. $=\frac{0.746}{4\pi}A^{4/3}\text{ e}^2\cdot\text{fm}^4$. Here, $A=A_\alpha+A_c=212$ is the mass number of $^{212}$Po, and $e$ is the elementary charge. The experimental values are taken from Ref.~\cite{Astier:2009bs,Astier:2010sn}.
%% For comparison, we also reproduce here the results from Ref.~\cite{Ibrahim:2010zz}.
}
\label{BE2}
\footnotesize
%\begin{center}
\begin{tabular}{cccc}
\hline
\hline
\hspace{0.5mm}Transition\hspace{0mm} & \hspace{0.5mm}{$B(\text{E}2\!\downarrow)_\text{Exp}$ (W.u.)}\hspace{0.5mm} & \hspace{0.5mm}{$B(\text{E}2\!\downarrow)_\text{Present}$ (W.u.)}\hspace{0.5mm}
% & \hspace{5mm}{$B(\text{E}2\downarrow)_\text{Ibrahim}$ (W.u.)}\hspace{5mm} 
\\[0.5ex]  
\hline
\hspace{5mm}$2^+\to0^+$\hspace{5mm} &  \hspace{5mm}{$-$}\hspace{5mm} & \hspace{5mm}{4.4}\hspace{5mm}
% & \hspace{5mm}{4.5}\hspace{5mm} 
\\
\hspace{5mm}$4^+\to2^+$\hspace{5mm} &  \hspace{5mm}{$-$}\hspace{5mm} & \hspace{5mm}{6.1}\hspace{5mm}
 %& \hspace{5mm}{6.2}\hspace{5mm}
  \\
\hspace{5mm}$6^+\to4^+$\hspace{5mm} &  \hspace{5mm}{$3.9\pm1.1$}\hspace{5mm} & \hspace{5mm}{6.3}\hspace{5mm}
% & \hspace{5mm}{6.4}\hspace{5mm} 
\\
\hspace{5mm}$8^+\to6^+$\hspace{5mm} &  \hspace{5mm}{$2.3\pm0.1$}\hspace{5mm} & \hspace{5mm}{5.9}\hspace{5mm}
% & \hspace{5mm}{6.0}\hspace{5mm} 
\\
\hspace{5mm}$10^+\to8^+$\hspace{5mm} &  \hspace{5mm}{$2.2\pm0.6$}\hspace{5mm} & \hspace{5mm}{5.2}\hspace{5mm}
% & \hspace{5mm}{5.3}\hspace{5mm} 
\\
\hspace{5mm}$12^+\to10^+$\hspace{5mm} &  \hspace{5mm}{$-$}\hspace{5mm} & \hspace{5mm}{4.4}\hspace{5mm}
% & \hspace{5mm}{4.4}\hspace{5mm} 
\\
\hspace{5mm}$14^+\to12^+$\hspace{5mm} &  \hspace{5mm}{$-$}\hspace{5mm} & \hspace{5mm}{3.4}\hspace{5mm}
% & \hspace{5mm}{3.4}\hspace{5mm} 
\\
\hspace{5mm}$16^+\to14^+$\hspace{5mm} &  \hspace{5mm}{$-$}\hspace{5mm} & \hspace{5mm}{2.3}\hspace{5mm}
% & \hspace{5mm}{2.3}\hspace{5mm} 
\\
 \hline
\hline
\end{tabular}
\end{center}
%\end{table}
%
%\begin{table}
\begin{center}
\tabcaption{Alpha-decay half-lives for different states of $^{212}$Po. $T_{1/2}^{\alpha,\text{Exp}}$ and $T_{1/2}^{\alpha,\text{Present}}$ denote the experimental and theoretical values of alpha-decay half-lives \cite{Astier:2009bs,Astier:2010sn}. The theoretical values are calculated with the performance factor $P_\alpha=0.6$.}
\label{AlphaDecayTable}
\footnotesize
%\begin{center}
\begin{tabular}{ccc}
\hline
\hline
\hspace{5mm}$J^\pi$\hspace{5mm} & \hspace{5mm}{$T_{1/2}^{\alpha,\text{Exp}}$ (ns)}\hspace{5mm} & \hspace{5mm}{$T_{1/2}^{\alpha,\text{Present}}$ (ns)}\hspace{5mm} \\[0.5ex]  
\hline
\hspace{5mm}$0^+$\hspace{5mm} &  \hspace{5mm}{299(2)}\hspace{5mm} & \hspace{5mm}{439}\hspace{5mm} \\
\hspace{5mm}$2^+$\hspace{5mm} &  \hspace{5mm}{$-$}\hspace{5mm} & \hspace{5mm}{16.31}\hspace{5mm} \\
\hspace{5mm}$4^+$\hspace{5mm} &  \hspace{5mm}{$-$}\hspace{5mm} & \hspace{5mm}{8.99}\hspace{5mm} \\
\hspace{5mm}$6^+$\hspace{5mm} &  \hspace{5mm}{25(10)}\hspace{5mm} & \hspace{5mm}{26.99}\hspace{5mm} \\
\hspace{5mm}$8^+$\hspace{5mm} &  \hspace{5mm}{490(150)}\hspace{5mm} & \hspace{5mm}{277}\hspace{5mm} \\
\hspace{5mm}$10^+$\hspace{5mm} &  \hspace{5mm}{$-$}\hspace{5mm} & \hspace{5mm}{$1.99\times10^3$}\hspace{5mm} \\
\hspace{5mm}$12^+$\hspace{5mm} &  \hspace{5mm}{$-$}\hspace{5mm} & \hspace{5mm}{$3.46\times10^3$}\hspace{5mm} \\
\hspace{5mm}$14^+$\hspace{5mm} &  \hspace{5mm}{$-$}\hspace{5mm} & \hspace{5mm}{$2.26\times10^5$}\hspace{5mm} \\
\hspace{5mm}$16^+$\hspace{5mm} &  \hspace{5mm}{$-$}\hspace{5mm} & \hspace{5mm}{$-$}\hspace{5mm} \\
\hspace{5mm}$18^+$\hspace{5mm} &  \hspace{5mm}{$4.5\times10^{10}$}\hspace{5mm} & \hspace{5mm}{$4.83\times10^{10}$}\hspace{5mm} \\
 \hline
\hline
\end{tabular}
\end{center}
%\end{table}
%
Given the above results on $^{212}$Po, we apply the WSG potential further to light nuclei $^{20}\text{Ne}={}^{16}\text{O}+\alpha$ and $^{44}\text{Ti}={}^{40}\text{Ca}+\alpha$. We take the same parameter set given by Eq.~\eqref{WSGParameters}, except $R=3.25$ fm for $^{20}$Ne and $R=4.61$ fm for $^{44}$Ti, which are determined by minimizing $\chi^2$ for the level schemes of $^{20}$Ne and $^{44}$Ti respectively. The quantum number $G$ is taken to be $G=8$ for $^{20}$Ne and $G=12$ for $^{44}$Ti. The calculated values are listed in Tables \ref{O16Table} and \ref{Ti44Table}, which also agree with the experimental values. Especially, the calculated values of electromagnetic transition strengths are in agreement with the experimental values within a factor of 1.1$\sim$1.3 for $^{20}$Ne and 1$\sim$1.7 for $^{44}$Ti.

\end{multicols}

%\begin{table}
\begin{center}
\tabcaption{The level scheme and electromagnetic transition strengths of $^{20}$Ne. The parameter set is given by Eq.~\eqref{WSGParameters} with a new value of $R=3.25$ fm. The experimental values are taken from Refs.~\cite{NNDC,Abele:1993zz}. 
Values for $B(\text{E}2\!\downarrow)$ are in the unit of the so-called Weisskopf unit (W.u.) given by 1 W.u. $=\frac{0.746}{4\pi}A^{4/3}\text{ e}^2\cdot\text{fm}^4$. Here, $A=A_\alpha+A_c=20$ is the mass number of $^{20}$Ne, and $e$ is the elementary charge.
}
\label{O16Table}
\footnotesize
%\begin{center}
\begin{tabular}{ccccc}
\hline
\hline
\hspace{4mm}$J^\pi$\hspace{4mm} & \hspace{4mm}{$E^{\text{Exp}}$ (MeV)}\hspace{4mm} & \hspace{4mm}{$E^{\text{Present}}$ (MeV)}\hspace{4mm} &  \hspace{4mm}{${B(\text{E}2\!\downarrow)_\text{Exp}}$ (W.u.)}\hspace{4mm} & \hspace{4mm}{${B(\text{E}2\!\downarrow)_\text{Present}}$ (W.u.)}\hspace{4mm} \\[0.5ex]  
\hline
\hspace{4mm}$0^+$\hspace{4mm} & \hspace{4mm}{0.000}\hspace{4mm} &  \hspace{4mm}{1.117}\hspace{4mm}  & \hspace{4mm}{$-$}\hspace{4mm} & \hspace{4mm}{$-$}\hspace{4mm} \\
\hspace{4mm}$2^+$\hspace{4mm} & \hspace{4mm}{$1.634$}\hspace{4mm} &  \hspace{4mm}{2.238}\hspace{4mm}  & \hspace{4mm}{$20.3\pm1.0$}\hspace{4mm} & \hspace{4mm}{18.341}\hspace{4mm} \\
\hspace{4mm}$4^+$\hspace{4mm} &  \hspace{4mm}{$4.248$}\hspace{4mm} & \hspace{4mm}{4.472}\hspace{4mm}  & \hspace{4mm}{$22.0\pm2.0$}\hspace{4mm} & \hspace{4mm}{23.663}\hspace{4mm} \\
\hspace{4mm}$6^+$\hspace{4mm} & \hspace{4mm}{8.776}\hspace{4mm} &  \hspace{4mm}{7.707}\hspace{4mm} &  \hspace{4mm}{$20.0\pm3.0$}\hspace{4mm} & \hspace{4mm}{19.314}\hspace{4mm} \\
\hspace{4mm}$8^+$\hspace{4mm} &  \hspace{4mm}{11.951}\hspace{4mm} & \hspace{4mm}{11.852}\hspace{4mm}  & \hspace{4mm}{$9.03\pm1.3$}\hspace{4mm} & \hspace{4mm}{9.867}\hspace{4mm} \\
 \hline
\hline
\end{tabular}
\end{center}
%\end{table}

%
%\begin{table}
\begin{center}
\tabcaption{The level scheme and electromagnetic transition strengths of $^{44}$Ti. The parameter set is given by Eq.~\eqref{WSGParameters} with a new value of $R=4.61$ fm. The experimental values are taken from Refs.~\cite{NNDC,Buck:1995zza}. 
Values for $B(\text{E}2\!\downarrow)$ are in the unit of the so-called Weisskopf unit (W.u.) given by 1 W.u. $=\frac{0.746}{4\pi}A^{4/3}\text{ e}^2\cdot\text{fm}^4$. Here, $A=A_\alpha+A_c=44$ is the mass number of $^{44}$Ti, and $e$ is the elementary charge.
}
\label{Ti44Table}
%\begin{center}
\begin{tabular}{ccccc}
\hline
\hline
\hspace{4mm}$J^\pi$\hspace{4mm} & \hspace{4mm}{$E^{\text{Exp}}$ (MeV)}\hspace{4mm} & \hspace{4mm}{$E^{\text{Present}}$ (MeV)}\hspace{4mm} &  \hspace{4mm}{${B(\text{E}2\!\downarrow)_\text{Exp}}$ (W.u.)}\hspace{4mm} & \hspace{4mm}{${B(\text{E}2\!\downarrow)_\text{Present}}$ (W.u.)}\hspace{4mm} \\[0.5ex]  
\hline
\hspace{4mm}$0^+$\hspace{4mm} &  \hspace{4mm}{0.000}\hspace{4mm} & \hspace{4mm}{0.695}\hspace{4mm} &  \hspace{4mm}{$-$}\hspace{4mm} & \hspace{4mm}{$-$}\hspace{4mm} \\
\hspace{4mm}$2^+$\hspace{4mm} &  \hspace{4mm}{$1.083$}\hspace{4mm}  & \hspace{4mm}{1.273}\hspace{4mm}  & \hspace{4mm}{$13.0\pm4.0$}\hspace{4mm} & \hspace{4mm}{13.152}\hspace{4mm} \\
\hspace{4mm}$4^+$\hspace{4mm} &  \hspace{4mm}{$2.454$}\hspace{4mm} & \hspace{4mm}{2.359}\hspace{4mm}  & \hspace{4mm}{$30.0\pm6.0$}\hspace{4mm} & \hspace{4mm}{17.651}\hspace{4mm} \\
\hspace{4mm}$6^+$\hspace{4mm} &  \hspace{4mm}{4.015}\hspace{4mm} & \hspace{4mm}{3.813}\hspace{4mm}  & \hspace{4mm}{$17.0\pm3.0$}\hspace{4mm} & \hspace{4mm}{16.881}\hspace{4mm} \\
\hspace{4mm}$8^+$\hspace{4mm} &  \hspace{4mm}{6.509}\hspace{4mm} & \hspace{4mm}{5.476}\hspace{4mm}  & \hspace{4mm}{$\!\!\!\!\!\!>1.52$}\hspace{4mm} & \hspace{4mm}{13.752}\hspace{4mm} \\
\hspace{4mm}$10^+$\hspace{4mm} &  \hspace{4mm}{7.671}\hspace{4mm} & \hspace{4mm}{7.152}\hspace{4mm}  & \hspace{4mm}{$\!\!\!\!\!\!14.97\pm3.0$}\hspace{4mm} & \hspace{4mm}{9.307}\hspace{4mm} \\
\hspace{4mm}$12^+$\hspace{4mm} & \hspace{4mm}{8.039}\hspace{4mm} & \hspace{4mm}{8.555}\hspace{4mm} & \hspace{4mm}{$\!\!\!\!\!\!<6.51$}\hspace{4mm}  & \hspace{4mm}{4.459}\hspace{4mm} \\
 \hline
\hline
\end{tabular}
\end{center}
%\end{table}

\vspace{5mm}

\begin{multicols}{2}

\section{Conclusions}

In summary, we study the alpha-cluster structures of various alpha+closed shell nuclei with a new phenomenological potential of the Woods-Saxon-Gaussian form, which modifies the original WS potential by an extra Gaussian form factor centered at the nuclear surface. %We would like to emphasize that it is by no means trivial to identify the radius in the Gaussian form factor with that in the WS potential.
% and might also be related to the recent study on an improved shell-model representation to describe alpha decays \cite{Delion:2013tla}. In that study, an extra Gaussian-shaped potential is added upon the ordinary shell-model mean field, which is also centered at the nuclear surface. It is quite natural to conjecture that these two Gaussian deformations to nuclear potentials might have the similar microscopic origins, such as the dissolution effect of the alpha cluster insides the core nucleus. 
%The WSG potential could be improved systematically by taking into consideration more Gaussian terms thanks to the over-completeness of the Gaussian basis. 
With the parameter set given by Eq.~\eqref{WSGParameters}, level schemes, reduced quadrupole transition strengths, and alpha-decay half-lives are found to be quite satisfying for $^{212}$Po, reproducing many important features of experimental values. This could be viewed as an explicit demonstration for the usefulness of our WSG potential in studying nuclear clustering in heavy nuclei. Moreover, we provide some exploratory calculations in studying alpha-cluster structures in light nuclei with the WSG potential. With the same parameter set Eq.~\eqref{WSGParameters}, except that new values are taken for the radius $R$, we calculate the level schemes and electromagnetic transitions for $^{20}$Ne and $^{44}$Ti. The calculated values agree with the experimental ones as well. These calculations show that the WSG potential proposed in this work is indeed a new phenomenological potential suitable for studying various properties of nuclear alpha-cluster structures. In this work, we have identified the radius in the Gaussian form factor with that in the WS potential. Considering the phenomenological success of the WSG potential, it is natural to ask for the microscopic origin of this choice, which might be related to the alpha-cluster formation process taking place at the nuclear surface and is an important open question to be investigated in future works.

\end{multicols}

\vspace{5mm}

\begin{multicols}{2}

\subsection*{Appendices A}
\begin{small}

\noindent{\bf Momentum-Space Approach to Double-Folding Potentials}

%\section{}
\label{MSA}
%In this appendix, we provide some analytic results on the double-folding potentials used in this note. 
The double-folding potential is given by
\begin{align}
&U(\mathbf{r})=\lambda\int\rho_c(\mathbf{r}_1)\rho_\alpha(\mathbf{r}_2)V_\text{NN}(\mathbf{r}_{12}=\mathbf{r}+\mathbf{r}_{2}-\mathbf{r}_{1})\mathrm{d}\mathbf{r}_1\mathrm{d}\mathbf{r}_2,\label{DFP}
%&V_\text{NN}(\mathbf{r}_{12})=7999\frac{\exp(-4r_{12})}{4r_{12}}-2134\frac{\exp(-2.5r_{12})}{2.5r_{12}}-276(1-0.005E_\alpha/A_\alpha)\delta(\mathbf{r}_{12}).
%\label{NNF}
\end{align} 
%The first two terms in Eq.~\eqref{NNF} are the Reid soft-core NN interaction, while the last term is the exchange component approximated by a zero-range pseudopotential.
with $\lambda$ being the renormalization factor. It is convenient to calculate double-folding potentials in the momentum space \cite{Satchler:1979ni}. The Fourier transformations are normalized by
\begin{align}
&\tilde{f}(\mathbf{k})=\int\mathrm{d}\mathbf{r}\exp(i\mathbf{k}\cdot\mathbf{r})f(\mathbf{r}),\\
&f(\mathbf{r})=\frac{1}{(2\pi)^3}\int\mathrm{d}\mathbf{k}\exp(-i\mathbf{k}\cdot\mathbf{r})\tilde{f}(\mathbf{k}),
\end{align}
which give $\int\mathrm{d}\mathbf{r}\exp(i\mathbf{k}\cdot\mathbf{r})=(2\pi)^3\delta(\mathbf{k})$. 
The density functions of the core nucleus and the alpha cluster and the nucleon-nucleon interaction take the following forms, respectively,
\begin{align}
&\rho_c(r_1)=\frac{\rho_1}{1+\exp\left(\frac{r_1-c}{a}\right)},
%=
%\begin{cases}
%\rho_1\{1+\sum_{n=1}^\infty(-1)^n\exp[-n(c-r_1)/a]\}, \qquad & r_1<c,\\
%\rho_1\sum_{n=1}^\infty(-1)^n\exp[-n(r_1-c)/a],\qquad & r_1>c,
%\end{cases}
\\
&\rho_\alpha(r_2)=\rho_2\exp(-\nu r_2^2),\\
&V_\text{NN}(\mathbf{r}_{12})=V_1\frac{\exp(-\mu_1r_{12})}{\mu_1r_{12}}+V_2\frac{\exp(-\mu_2r_{12})}{\mu_2r_{12}}+J_{00}\delta(\mathbf{r}_{12}),
\end{align}
with their Fourier transformations given by
\begin{align}
&\tilde{\rho}_c(k)=4\pi\int_0^{\infty}\mathrm{d}r_1\, r_1 \frac{\sin kr_1}{k} \rho_c(r_1),\\
%=\sum_{n=1}^{\infty}\frac{4\pi\rho_1\exp(-nc/a)}{k^3(a^2k^2+n^2)^2}\Big\{-2n(-1)^na^3k^3+\exp(nc/a)\Big[-(a^2k^2+n^2)^2(ck\cos ck-\sin ck)\nonumber\\
%&\qquad\ +2n(-1)^nak^2(2a^2k\cos ck+c(a^2k^2+n^2)\sin ck)\Big]\Big\},\\
&\tilde{\rho}_\alpha(k)=\rho_2\pi^{3/2}\nu^{-3/2}\exp\left(-\frac{k^2}{4\nu}\right),\\
&\tilde{V}_\text{NN}(k)=\frac{4\pi V_1}{\mu_1}\frac{1}{k^2+\mu_1^2}+\frac{4\pi V_2}{\mu_2}\frac{1}{k^2+\mu_2^2}+J_{00}.
\end{align}
%It is easy to check that $\tilde{\rho}_c(k)$, $\tilde{\rho}_\alpha(k)$, and $\tilde{V}_\text{NN}(k)$ are all even functions of $k$. 
The double-folding potential could then be obtained by
\begin{align}
U(\mathbf{r})&=\frac{1}{(2\pi)^3}\int\mathrm{d}\mathbf{k}\exp(-i\mathbf{k}\cdot\mathbf{r})\tilde{\rho}_c(k)\tilde{\rho}_\alpha(k)\tilde{V}_\text{NN}(k)\\
&=\frac{2}{\pi}\int_{0}^\infty\int_{0}^\infty\mathrm{d}k\,\mathrm{d}r_1 \frac{r_1}{r}{\sin kr}\,{\sin kr_1} {\rho}_c(r_1)\tilde{\rho}_\alpha(k)\tilde{V}_\text{NN}(k).\label{2DDF}
%&=-\frac{i}{4\pi^2r}\oint_\mathcal{C}\mathrm{d}k\, k \exp(ikr) \tilde{\rho}_c(k)\tilde{\rho}_\alpha(k)\tilde{V}_\text{NN}(k)\\
%&=\frac{1}{2\pi r} \left\{\text{Res}[u(k),i\mu_1]+\text{Res}[u(k),i\mu_2]+\sum_{n=1}^{\infty}\text{Res}[u(k),ina^{-1}]\right\}.
\end{align}
Compared with the definition of the double-folding potential Eq.~\eqref{DFP} which contains a six-dimensional integration, Eq.~\eqref{2DDF} with only a two-dimensional integration is much easier to handle numerically.
The parameter set used in this work is given by
\begin{align}
&\rho_1=0.181855\text{ fm$^{-3}$},\quad c=1.07A^{1/3}_c\text{ fm},\quad a= 0.54 \text{ fm},\nonumber\\
&\rho_2=0.4299\text{ fm}^{-3},\quad \nu=0.7024\text{ fm}^{-2},\nonumber\\
&V_1=7999 \text{ MeV},\quad V_2=-2134\text{ MeV}, \nonumber\\
&J_{00}=-276(1-0.005E_\alpha/A_\alpha) \text{ MeV}\cdot\text{fm}^3 ,\nonumber\\
&\mu_1=4 \text{ fm}^{-1},\quad \mu_2=2.5 \text{ fm}^{-1},\quad \lambda=0.51,
\end{align}
with $A_\alpha=4$, $A_c=208$, and $E_\alpha$ being the alpha-decay energy.

\end{small}

\end{multicols}

\vspace{-1mm}
\centerline{\rule{80mm}{0.1pt}}
\vspace{2mm}

\begin{multicols}{2}

\end{multicols}

\clearpage

%\end{CJK*}
\end{document}